\documentclass[aps,twocolumn,pre,showpacs,floatfix,superscriptaddress]{revtex4}
\usepackage{graphicx,amsmath,epsfig}
\usepackage{amssymb}
\usepackage{bm}

\def\be{\begin{equation}}
\def\ee{\end{equation}}
\def\bea{\begin{eqnarray}}
\def\eea{\end{eqnarray}}
\def\la{\label}
\def\bsea{\begin{subeqnarray}}
\def\esea{\end{subeqnarray}}

\begin{document}

\title[]{ Coarsening of two
        dimensional $XY$ model with Hamiltonian dynamics: Logarithmically divergent vortex mobility}


\author{Keekwon \surname{Nam}}
\affiliation{Department of Physics, Changwon National University, Changwon 641-773, Korea}

\author{Woon-Bo \surname{Baek}}
\affiliation{Department of Mechatronics, Dong-Eui University, Busan 614-714, Korea}

\author{Bongsoo \surname{Kim}}
\affiliation{Department of Physics,
Changwon National University, Changwon 641-773, Korea}

\author{Sung Jong \surname{Lee}}
\affiliation{Department of Physics, University of Suwon, Hwaseong-Si, 445-743, Korea}

\date{\today}

\begin{abstract}

We investigate the coarsening kinetics of an $XY$ model defined on a square lattice
when the underlying dynamics is governed by energy-conserving Hamiltonian equation of motion.
We find that the apparent super-diffusive growth of the length scale can be interpreted as the vortex 
mobility diverging logarithmically in the size of the vortex-antivortex pair, 
where the time dependence of the characteristic length scale can be fitted as 
$L(t) \sim  ((t+t_{0}) \ln(t+t_{0}))^{1/2}$ with a finite offset time $t_0$. 
This interpretation is based on a simple phenomenological model of vortex-antivortex annihilation to 
explain the growth of the coarsening length scale $L(t)$. 
The nonequilibrium spin autocorrelation function $A(t)$ and the growing length
scale $L(t)$ are related by $A(t) \simeq L^{-\lambda}(t)$ with a distinctive exponent 
of $\lambda \simeq 2.21$ (for $E=0.4$) possibly reflecting the strong effect of propagating 
spin wave modes.   
We also investigate the nonequilibrium relaxation (NER) of the system under
sudden heating of the system from a perfectly ordered state to the regime of
quasi-long-range order, which provides a very accurate estimation of the 
equilibrium correlation exponent $\eta (E) $ for a given energy $E$.  
We find that both the equal-time spatial correlation $C_{nr}(r,t)$ and the 
NER autocorrelation $A_{nr}(t)$ exhibit scaling features consistent with the dynamic
exponent of $z_{nr} = 1$.      

\end{abstract}

\pacs{64.60.Ht,64.60.Cn,75.10.Hk,75.40.Gb}
\keywords{Hamiltonian dynamics, Coarsening, Phase ordering, Dynamic scaling, $XY$ model}

\maketitle

\section{INTRODUCTION}

Phase ordering kinetics refers to a study of nonequilibrium dynamic processes of 
statistical mechanical systems approaching toward the thermal equilibrium of ordered 
states with broken symmetry. The system is usually quenched from a random disordered 
state to a low-temperature ordered state below the transition 
temperature \cite{ordering_review,bray_review}. The ever-nonequilibrium approach to 
ordered state, in the thermodynamic limit, is usually characterized by the dynamic 
scaling accompanied by power-law growth of typical length scales of the system. 
These power law exponents depend on the spatial dimension and the symmetry of the ground 
states. More importantly these power laws also depend on the nature of the dynamic 
protocols such as whether the order parameter is conserved or not during the phase 
ordering processes.
Usually, characteristic topological defects, such as point vortices or domain
walls, are generated in the initial disordered state, depending on the dimension 
of the order parameter, and the phase ordering processes can also be interpreted as the 
process of annihilation and decay of these topological defects: the motion of these 
defects (and their annihilation) determines the characteristics of the coarsening 
processes. 

Most of the computational works on phase ordering kinetics have been based on a purely 
dissipative dynamics. In this case, we expect that the motion of the vortices would 
be also purely dissipative. 
Energy scaling arguments in combination with conservation laws could determine 
the growth laws for most of the $O(n)$ spin systems in $d$ dimensions \cite{bray_review} 
(except for the special case of $d=n=2$). 
       
In reality, however, there exist various systems exhibiting dynamic processes that 
cannot be described solely by dissipative dynamics. 
For example, inertial effects would be present in the case of magnetic spin systems, where 
the spins are influenced by neighboring spins via precession interaction terms that are energy
conserving \cite{ma_mazenko, nelson-fisher, Nam_mcxy} that are often called reversible mode coupling. 
Actually, among the eight model systems classified by Hohenberg and 
Halperin \cite{hohen77,folk2006} 
for describing the dynamic critical phenomena, five of them (models E, F, G, H, and J) 
include reversible mode couplings. 

We may also take a microcanonical approach to the nonequilirium dynamics by
modeling the dynamics of the system based on a purely Hamiltonian dynamics with
energy conservation. In relation to this, there have been some research efforts in recent years
toward investigating the statistical mechanical behavior of lattice spin systems by solving 
directly the Hamiltonian equation of motion associated with the spin system \cite{casetti_hamil_review}. 
For example, equilibrium phase transitions have been found 
to be related to a topological change in the 
dynamics \cite{caian1, caian2,leoncini,ruffo2001, cerruti, latora}.
In terms of nonequilibrium phase ordering dynamics, Zheng investigated via Hamiltonian dynamics
the coarsening of the Ising-type $\phi^4$ model on a two-dimensional square lattice with 
nonconserved order parameter \cite{zheng_ising_ordering, kockel_ising_ordering_comment}.

In this work, we study numerically the coarsening process of an $XY$ model on 
a square lattice using Hamiltonian dynamics. This work is built upon a previous preliminary 
work \cite{koo2006} on the same model system, which we think is incomplete,
especially in terms of the analysis on the growth law. 

In equilibrium, the  $XY$ model exhibits a Berezinskii-Kosterlitz-Thouless (BKT)
transition at $T_{BKT}$ due to the unbinding of vortex-antivortex pairs \cite{BKT}. 
Below $T_{BKT}$, the system has a quasi-ordered phase that is characterized by 
a power-law decay of the order parameter correlation function for long 
distances. The critical exponent governing the power-law decay decreases 
{\it continuously} down to zero temperature: the system is critical at 
equilibrium for all non-zero temperatures below $T_{BKT}$.

The phase ordering dynamics of the {\em XY} model has been studied for quite a long time by 
many groups \cite{loft,toyoki3,mondello,bray, yurke,blundell,jrl,rojas,bray2,bray3,ying}. 
It is now agreed that, in the phase ordering dynamics of the $XY$ model with 
non-conserving order parameter (via Monte Carlo simulations or Langevin dynamics methods), 
the growing length scale $L_{MC}(t)$ exhibits a logarithmic correction \cite{wilczek} to 
diffusive growth as $L_{MC}(t) \sim \big( t/\log t \big)^{1/2}$.  
Since the coarsening in $XY$ model is dominated by the annihilation of vortex-antivortex
pairs generated in the random initial states, we may interpret the growth law in terms 
of the mobility (or friction) characteristics of the vortices. 
Here, the logarithmic correction can be attributed to a logarithmic divergence 
(in the system size) of the effective friction constant of a moving vortex in the 
dissipative dynamics with a non-conserving order parameter. 
Another system that can be described by $XY$ model at equilibrium is an array of 
superconducting Josephson junctions. 
In the cases of superconducting-normal-superconducting (SNS) junction arrays, 
the dynamics is often described by resistively-shunted junction (RSJ) model.
This model is characterized by a Laplacian type of coupling between time derivatives
of superconducting phases of  neighboring islands.
Due to this particular type of dissipative coupling in Josephson junction arrays,
there is {\it no} logarithmic correction in the growing length scale that corresponds to  
a finite friction constant of a moving vortex \cite{jj_ordering}.

One further experimental system that can be described by $XY$ model is a thin film 
of superfluid helium. The dynamics of this system is characterized by a reversible 
mode-coupling term. In recent simulations on the coarsening dynamics of $XY$ model
with reversible mode-coupling (MCXY), it was shown that the growth law exhibits a 
{\it positive} logarithmic correction which is opposite to the coarsening of 
ordinary $XY$ model with purely dissipative dynamics and non-conserving order 
parameter \cite{Nam_mcxy}.
This system apparently exhibits a growth law with an exponent that is a little larger
than the diffusive growth exponent of $1/2$. But this was re-analyzed and fitted
in terms of a (positive) logarithmic correction to a diffusive growth. The growth law could 
also be understood in terms of a simple vortex-antivortex annihilation model. These 
features reminded us of the growth law with apparent super-diffusive exponents 
 in the coarsening of $XY$ model with Hamiltonian dynamics \cite{koo2006}.     
  
Now, in this work, we perform more extensive simulations on the Hamiltonian coarsening
of $XY$ model and carry out an analysis of the growth law in more details in order 
to determine whether these similarities between MCXY and Hamiltonian $XY$ models
are indeed valid. Due to the Hamiltonian dynamics nature of our system, the total 
energy is conserved, and we specify the initial states by their total energy divided by 
the system size, i.e., the per-site energy $E$. These initial states with specified energies 
are prepared with a special Monte Carlo algorithm \cite{koo2006}. 
The initial rotational velocities of the rotors are taken to be zeros, ensuring that 
the initial configuration is maximally disordered within the constraint of the fixed 
energy.  

Since we begin with zero kinetic energy (zero rotational velocities for all rotors), 
the Hamiltonian dynamics of the system will generate kinetic energy taken from 
the potential energy. As time proceeds, we can expect that (for a per-site energy 
that is low enough to correspond to a low-temperature quasi-ordered phase), 
the system will evolve toward some equilibrium stationary state which can be 
considered as corresponding to the ordinary thermal equilibrium state with critical
quasi-long-range order. 
In other words, the kinetic energy that is generated during the course of coarsening 
process acts as a thermal bath for the system. We investigate the time-dependent spin 
configuration for the system in terms of spatial ordering and relaxation of the vortex 
numbers, etc., in analogy to the conventional dissipative coarsening systems.

We find that the equal-time spatial correlation functions satisfy
critical dynamic scaling
\begin{equation} \label{eq:dynamic_scaling}
C(r,t)= r^{-\eta(E)} g(r/ L(t)),
\end{equation}
with a spatial correlation exponent $\eta (E)$ (which is increasing in 
the per-site energy $E$) and 
a growing length scale $L(t)$ that grows typically as $L(t) \sim t^{1/z} $, where 
$z$ is the dynamic exponent. In the late time region, we find that the length 
scale $L(t)$ grows with an exponent $1/z$ that is apparently a little larger than the 
diffusive exponent $1/2$. However, from more detailed analysis we could confirm 
that the growth of the length scale can be interpreted as the vortex mobility 
diverging logarithmically in the size of the vortex-antivortex pair. The growth of 
characteristic length scale can be fitted as $L(t) \sim  ((t+t_{0}) \ln(t+t_{0}))^{1/2}$ 
with a finite offset time $t_0$. This offset time $t_0$ corresponds to a finite 
length scale $L_0$ at $t=0$ due to the finite average spacing between vortices and
antivortices at the initial time.  

We could also compare the nonequilibrium spin autocorrelation function $A(t)$ of the
order parameter in the coarsening process and the growing length scale $L(t)$. 
Here, we found that $A(t)$ and the $L(t)$ are related by a scaling exponent $\lambda$ with 
$A(t) \simeq L^{-\lambda}(t)$ where $\lambda \simeq 2.21$ (at $E=0.4$) with only very 
weak monotonic increase on the value of $E$. This is rather close to that in
the case of MCXY or (soft-spin) model E \cite{Nam_mcxy}. 
While, on the other hand, this value of $\lambda$ exponent is quite different from that 
of usual dissipative coarsening dynamics of XY model (or $O(2)$ model) with nonconserved 
order parameter where the exponent of $\lambda \sim 1.17$ was predicted theoretically and
also observed in numerical simulations \cite{jrl, newman90,bh,newman90_2,lm92}.
This implies that the system loses its memory much
more quickly in the Hamiltonian dynamics or in the case of MCXY than
in the case of simple dissipative dynamics.       

Our simulation results indicate that the phase ordering kinetics of Hamiltonian $XY$ model
belongs to the same universality class as that of the Langevin dynamics of $XY$ model 
with a reversible mode-coupling and also that of (soft-spin) model E dynamics. This seems 
to be due to the existence of the same symmetry and conservation law in the two systems. 
That is, the total rotational momentum is conserved in the Hamiltonian dynamics which 
corresponds to the conservation of the third component (or the conjugate momentum) in the 
model E dynamics. Due to the conservation of the rotational momentum, propagating spin 
wave modes appear \cite{das-rao1, das-rao2}. 
This propagating spin wave modes might interact with the vortices and antivortices
such that the mobility diverges logarithmically as the distance between vortices  
and antivortices increases. As in the previous work on the phase ordering dynamics of 
MCXY model, here also we attempt to fit the growth of the length scale with a
simple phenomenological dynamic model of coarsening based on vortex-antivortex annihilation.    

In order to analyze the critical dynamic scaling of the equal-time spatial correlation
of the order parameter, we need an accurate estimate of the spatial correlation exponent
$\eta(E)$. For accurate measurement of $\eta(E)$, we employ an analog of the so-called
nonequilibrium relaxation (NER) method. 
Here one starts with a perfectly ordered initial state but with random finite rotational 
velocities such that the total kinetic energy per site is equal to $E$. 
Beginning with this initial state, the system evolves toward a steady dynamic state 
corresponding to a critical equilibrium state.
By analyzing the approach to the steady state in terms of the equal-time spatial correlation 
function of the order parameter, we can estimate more accurately the equilibrium 
correlation exponent $\eta (E)$. In this case of NER relaxation, we can also obtain
approximate analytic expression for the equal-time spatial correlation and autocorrelation
function by using spin-wave approximation (see APPENDIX A), which agrees qualitatively 
well with simulation results.

\bigskip

\section{The Hamiltonian XY model and Simulation Methods}

\bigskip

The dynamic equation of the hard-spin {\em XY} model on a square lattice 
(of linear size $N$) can be obtained from the following Hamiltonian given by
\begin{equation} \label{Hamiltonian}
H  = \frac{1}{2} \sum_i  m_i^2 + J \sum_{<ij>} \Big[ 1- \cos(\theta_i - \theta_j) \Big], 
\end{equation}
where, $J$ is the interaction strength, $\theta_i$ is the phase angle
of the spins at site $i$ and the sum is over nearest neighbor pairs. 
In the kinetic term, $m_i = \dot{\theta}_i$ represents the angular momentum associated 
with the angular variable $\theta_i$ of planar spin at site $i$. This 
model appears naturally in easy-plane ferromagnets or superfluid helium 
with the conserved variable $m$ corresponding to the $z$-component of 
the spin or the density of the superfluid in the limit of negligible thermal 
noise (e.g., negligible phonon effect) \cite{hohen77,nelson-fisher}.             
  
Previous works \cite{leoncini, ruffo2001} have shown that a $BKT$ transition occurs 
at the value of the per-site energy around $E_{BKT}\simeq 1.0$. 
In this work, we perform simulations of Newtonian dynamics based on the above Hamiltonian 
with initial states of zero kinetic energy and a pre-specified potential energy $E$ 
(but otherwise with random initial phases). Specified potential energy values are 
chosen from the region that corresponds to below the $BKT$ transition 
energy ($E_{BKT} \simeq 1.0$).

Unlike the case of dissipative systems with Langevin dynamics or 
Monte Carlo dynamics where dissipative relaxation is incorporated 
in a natural way, here in our case of Hamiltonian dynamics, the 
total energy (sum of kinetic and potential energy) is conserved. 
However, since we begin the Hamiltonian dynamics with zero kinetic energy 
with a given specified maximal potential energy, we expect that the system will 
evolve in such a way as to decrease (on the average) the potential energy  
and increase the kinetic energy.  Eventually, in the long-time limit, 
the system will reach a certain kind of near-equilibrium dynamic state where 
the average potential energy and the kinetic energy do not vary appreciably. 
Of course, rigorously speaking, in the thermodynamic limit, the system will 
never reach a complete equilibrium state in a finite time period.

Therefore, we can consider the kinetic part of the system acting as
a kind of thermal heat bath that absorbs the "dissipated heat"
coming from the decreasing potential energy. We can, thus, investigate
the time-dependent spin configurations for the system in terms of
spatial ordering and relaxation of vortex numbers, etc. 
We generate the
initial states with specified potential energies in the following
manner. To begin with, we take a random spin configuration. Then we
employ a Monte Carlo annealing algorithm such that the spin
configurations with potential energies that are closer to the target
energy are preferentially accepted. In other words, we steer the
initial state to some target spin configuration such that we obtain
a state whose potential energy is equal to a specific energy. 
In other respects (i.e., except for the constraint of the specified energy), 
the spin configuration is kept random.

The Hamiltonian equation for Eq.~(\ref{Hamiltonian}) reads
\begin{eqnarray} \label{eq:hxy-eq}
{\dot{m}_{i} } & = & - \frac{\partial U}{\partial \theta_{i}} = -J \sum_{j} 
\sin (\theta_i -\theta_{j}), \\
{\dot{\theta}_{i}}   &    =  & m_{i}
\end{eqnarray}
where $j$ denotes the nearest neighbors of site $i$ with $i=1, \cdots, N^2$. 
For simplicity, we set the rotational inertia to be equal 
to unity and put $J=1$. Here, we note that, in addition to the conservation 
of total energy, another conserved quantity exists, namely the total rotational 
momentum $\sum_{i} m_{i} \equiv \sum_{i}\dot{\theta}_{i}$. 
This is due to the invariance of the equations under global rotations 
in the variable $\theta_{i}$, that is, under 
$\theta_{i} \rightarrow \theta_{i} + \alpha$.  

Equations~(\ref{Hamiltonian}) and (\ref{eq:hxy-eq}) are numerically integrated in time 
using a second order velocity-Verlet algorithm \cite{yoshida} with the time 
integration step of $\Delta t = 0.01$, which conserves the total energy to 
within a ratio of $10^{-4}$ up to the maximal time integration steps of around 
$10^6$. Periodic boundary conditions on both lattice directions are employed. 
The simulations are carried out on square lattices of dimensions up to 
$N\times N =2000 \times 2000$ with sample averages of over $40$ to $150$ 
different initial configurations. A parallel computation is employed via a domain 
decomposition method with high parallel efficiency.

The main quantities of interest are as follows:

(i) Equal-time spatial correlation function of the order parameter,

\begin{equation}
C (r,t)  =  \frac{1}{N^2} \left < \sum_{i} \cos (\theta_{i}(t) - \theta_{i+r}(t) \right >,
\end{equation}
where $< \cdots >$ denotes an average over random initial configurations.

(ii) The total number of vortices and antivortices $N_V (t)$ at time $t$, 
     and 

%
%

(iii) Nonequilibrium spin autocorrelation function

\begin{equation}
{A} (t) \equiv \frac{1}{N^2} \left < \sum_{i}
\cos ( \theta_{i}(0) - \theta_{i}(t) ) \right >.
\end{equation}


\bigskip

\noindent

\section{Simulation Results and Discussions}

\bigskip


\subsection{Nonequilibrium relaxation (NER) from fully ordered state}

As was mentioned in the introduction, the equal-time spatial correlation function in
the coarsening process is expected to obey a critical dynamic scaling 
(Eq.~(\ref{eq:dynamic_scaling})). One of the parameters to be determined is the 
equilibrium correlation exponent $\eta (E) $. We may try to fit the equal-time spatial
correlation function for the coarsening dynamics with $\eta (E)$ as a free parameter.
In this case, various values of $\eta (E) $ should be chosen on a trial and error basis,
and the best dynamic scaling collapse must be chosen among these many possibilities.
This is not a very accurate or reliable procedure let alone being tedious.

Instead, we have a better way to estimate the values of $\eta(E)$ independently.        
This is the so-called NER method where, in the usual case, the system is instatntaneously brought 
from a perfectly ordered state to a finite temperature $T$. This target temperature is 
usually chosen as the critical temperature corresponding to a second order phase transition. 
Here, in our case of Hamiltonian dynamics, the energy per site $E$ is conserved. 
It is easy to see that an analog of NER method here can be achieved as follows.
Similar to the case of usual lattice spin system under thermal noise, here, one 
starts with a perfectly ordered initial state. But we have to remember that the per-site energy 
of the system has to be conserved (equal to $E$). Since the perfectly ordered initial
state has zero potential energy (from our definition of the Hamiltonian in Eq.~(\ref{Hamiltonian})),  
the rotational velocities $\dot{\theta}_i $ in the initial configuration should be chosen
such that the initial per-site kinetic energy be equal to $E$. Except for this constraint 
of energy conservation, the initial velocities are chosen randomly.
Beginning with this initial state of perfect order but with finite kinetic energy (through
random rotational velocities), when the initial kinetic energy per site is low enough, the 
system will evolve toward a steady state corresponding to a critical equilibrium 
state at a finite temperature (correspondng to temperatures below the $BKT$ transition).      

By analyzing the approach to the steady state in terms of the equal-time spatial correlation 
function of the order parameter, we can compute more accurately the equilibrium correlation 
exponent $\eta (E)$.     
The NER equal-time spatial correlations for the case of $E=0.4$ are shown in Fig.~1a. As time
elapses, spatial extent of critical power law relaxation increases with a clear power exponent.
In order to extract the $\eta$ exponent more rigorously, we analyze the data with a critical 
dynamic scaling
\begin{equation} \label{eq:NER_scaling}
C_{nr}(r,t)= r^{-\eta(E)} g_{nr}(r/ L_{nr}(t)),
\end{equation}
with the spatial correlation exponent $\eta (E)$ and a growing correlation length scale 
$L_{nr}(t)$ with $L_{nr}(t) \sim t^{1/z_{nr}}$.
A good collapse of the rescaled correlation functions is shown in Fig.~1b. 
Interestingly, we find that the correlation length scale $L_{nr}(t)$ exhibits a linear 
growth in time as $L_{nr}(t) \sim t $ i.e., $1/z_{nr} \simeq 1$ which is quite different 
from the case of nonconserved dissipative dynamics where the diffusive exponent $1/2$ is 
observed \cite{kknam_hxy_unpub}. This may be attributed to the effect of propagating spin wave mode
in the linearized equation of motion.  Indeed, in the spin-wave approximation, we find that 
the equal-time spatial correlation $C_{nr}(r,t)$ can be expressed in closed form (see the Appendix A) 
which exhibits critical dynamic scaling and linear growth of the length scale $L_{nr}(t)$
with 
\begin{equation} \label{eq:NER_scaling_SW}
C_{nr:SW}(r,t) \simeq r^{-\eta_{SW} (E)} g_{nr:SW}(r/2t))
\end{equation}
where $\eta_{SW} (E) = E / 2 \pi $ and
\begin{equation} \label{eq:scaling_SW}
\begin{array}{rcl}
g_{nr:SW}(u) & =  &  u^{\eta_{SW} (E)} 
\left (\displaystyle\frac{1 + \sqrt{1-u^2}}{ 1 - \sqrt{1-u^2}} \right )^{\eta_{SW} (E)/2 } \hspace{0.2cm} (u<1), \\
      &      &          \\
     & = &  u^{\eta_{SW} (E)} \hspace{0.5cm} (u > 1). 
\end{array}
\end{equation}
where $SW$ in the subscript refers to ``spin-wave approximation".
The behavior of $C_{nr}(r,t)$ vs. $r$ at the latest time 
$t=320$ in simulation for several cases of the energy $E$ are shown in Fig.~1c with the 
$\eta(E)$ exponents and the spin-wave approximation values $\eta_{SW}(E)$ in the inset. 
We can see an approximate agreement between the simulation results of $\eta(E)$ and $\eta_{SW}(E)$
especially at low $E$ which is consistent with the spin-wave approximation.
If we consider the case of $E=0.4$, the simulation results of $\eta (E=0.4)=0.071$ can be compared 
reasonably with the approximate analytic value of $\eta_{SW} (E) = E/2 \pi \simeq  0.064 $ at 
spin-wave approximation. Also, with the same approximation, we could easily derive the formula
for the autocorrelation $A_{nr:SW}(t)$ as $ A_{nr:SW}(t) \sim t^{-\eta_{SW} (E)/2} $ (see APPENDIX A). 

\subsection{Critical coarsening dynamics}

Now we turn to the critical coarsening dynamics evolving from random configurations.     
Figure~2a shows the equal-time spatial correlations for $E=0.4$ for different time instants.
We attempt a critical dynamic scaling
\begin{equation} \label{eq:dynamic_scaling2}
C(r,t)= r^{-\eta(E)} g(r/ L(t)),
\end{equation}
where we use the spatial correlation exponent $\eta (E)$ obtained from the NER analysis 
shown above. In our case, for a given time instant $t$, we determined $L(t)$ in such a 
way that $ r^{\eta(E)} C(r,t)|_{(r=L(t))} = g(r/L(t))|_{(r=L(t))} = g(1) = 0.4 $.
Shown in Fig.~2b is such an attempt at critical dynamic scaling where we find that the
equal-time spatial correlation functions exhibit a reasonable critical dynamic scaling, 
at least, in the late-time regime. 

In the late-time regime, we find that the length scale $L(t)$ grows
with an effective exponent $1/z$ that is apparently larger than the diffusive
exponent $0.5$. For example, in the case of $E=0.4$, we obtain $1/z \simeq 0.553 \pm 0.005$, 
which is a little larger than $1/2$ (Fig.~2c). This is consistent with the previous work \cite{koo2006}. 
In contrast, in the conventional purely diffusive case, the effective growth exponents 
obtained numerically are invariably smaller than $1/2$ due to a negative logarithmic 
correction. 
Similar enhancement of growth was observed recently in the coarsening dynamics of 
the (soft spin) model E as well as the MCXY model.
There, it was interpreted as the result of a logarithmically divergent mobility of vortices 
(and antivortices) in terms of a simplified kinetic model of a single pair of vortex and antivortex.     
Here we find that the same model can explain the coarsening process only with a modification
of the initial condition that accounts for the finite length scale $L_0$ at time $t=0$ due
to the special nature of Hamiltonian dynamics and corresponding energy conservation in 
the present system.         
Similar phenomenon of logarithmically divergent mobility was also found experimentally in the
quasi-two-dimensional diffusion of colloids \cite{sane_2009} or diffusion of protein
molecules on the membranes \cite{saffman_1975} under hydrodynamic effect, and also in 
numerical simulations of the vortex diffusion in the anisotropic Heisenberg system in two 
dimensions with spin precession \cite{bishop, kamppeter1, kamppeter2}.

We reiterate briefly on the single vortex-antivortex annihilation model.
We can see that, in order for the system to reach a state with average length scale of $R$, 
vortex pairs of sizes on the order of $R$ must be already annihilated. 
Ignoring the very complicated many-body process involving many vortex pairs during this process,
we simplify the whole coarsening process (corresponding to the growth of the length scale
up to $R$) by the annihilation of {\em single} vortex-antivortex pair of size $R$ with
a suitably defined interaction potential.

We assume that the vortex acts like a small particle with finite 'mass'   moving under 
the influence of an external force with a mobility that depends logarithmically on the 
length scale of the system. Then the distance $R$ between a vortex and an anti-vortex 
would be described by the following equation of motion
\begin{equation} \label{eq:vort_anni_model}
m(R) { \frac{d^2 R}{dt^2} } +  \frac{1}{\mu (R)} {\frac{dR}{dt} } =F(R) =  - \frac{k}{R}.
\end{equation}
Here, the length-dependent effective mass of a vortex is denoted as $m(R)$ and
the length-dependent mobility of a vortex as $\mu(R)$. We also assume that
the vortex-antivortex pair is interacting via a Coulombic force $F(R) =-k/R$ in two
dimensions. We will set the proportionality constant $k$ as $k=1$.
As for the functional form of the $m(R)$ and $\mu (R)$,  we use the same forms  
as that employed in \cite{Nam_mcxy} where logarithmic dependences of the
vortex mobility $\mu(R)$ and $m(R)$ are used \cite{bishop, kamppeter1, kamppeter2} as
follows
\begin{equation} \label{eq:length_dep_mass}
m(R) = m_0 + m_1 \ln (R/r_0), \qquad  \mu (R) = \mu_0 + \mu_1 \ln (R/r_0)
\end{equation}
where $m_0$, $m_1$, $\mu_0$, $\mu_1$ are constants and $r_0$ denotes a
shortest cutoff length scale in the system (corresponding to the vortex core size).
Note that $m_0$ corresponds to the effective mass at the shortest cutoff length
scale and similarly for $\mu_0$ for the vortex mobility.

The growth law is obtained by calculating (through integration of the above model 
dynamic equation) the time $\tau$ it takes for a vortex-antivortex pair of size 
$R = R^*$ to reach a state with $R=R_0$ where $R_0$ denotes the initial (coarsening) 
length scale of the system (the average separation between vortices or antivortices). 
We set the value of $dR/dt$ at $R= R^*$ to be zero. 
We numerically solve the model equation of vortex-antivortex pair dynamics, 
Eqs.~(\ref{eq:vort_anni_model}) and (\ref{eq:length_dep_mass}), to obtain the time $\tau$  when 
the size of the vortex-antivortex 
pair $R$ becomes equal to $R_0$. Note that $R_0$ is larger than the lower cutoff length scale of 
$r_0$ corresponding to the unit lattice spacing. 
By plotting the resulting relation $R^*$ and $\tau$ we get the growth law of 
the coarsening dynamics.

If we assume that the inertial effect is negligible, then, due to the logarithmic 
divergence of the vortex mobility, we can analytically get the dominant asymptotic 
growth law as
\begin{equation} \label{growth_law_log}
L(t) \sim \big( t \ln t \big)^{1/2}.
\end{equation}
However, we note that, in the limit of $t\rightarrow 0$, the domain length scale $L_0$
obtained from our simulations of our Hamiltonian $XY$ model approaches a finite 
value which is incompatible with the above form (which gives $L(t=0) = 0$). 
In order to reconcile with this limiting conditions, we should take a more general
ansatz equation for the time dependence of the length scale that exhibits a finite initial 
value for the length scale and still satisfies the dynamic equation of motion asymptotically.               
One plausible candidate is the following form which is obtained by a simple shift of 
the time variable $t$ as
\begin{equation} \la{growth_law_log_shift}
L(t) \sim \big( (t+t_0 ) \ln (t+t_0 ) \big)^{1/2}.
\end{equation}
We might also take another form by adding a constant $L_0$ to Eq.~(\ref{growth_law_log}) as
$ L(t) \sim  L_0 +  a (t \ln t)^{1/2} $. But here we chose the above method of shifting
the time because it appears the most natural, being analytic at $t=0$.
We find that $L(t)$ vs. $t$ can be fitted by Eq.~(\ref{growth_law_log_shift}) in the late time regime (Fig.~2c). 

In addition, shown in Fig.~3a is the relaxation of the vortex number density $\rho_v (t)$ 
for $E=0.4$. It exhibits a power law relaxation with $\rho_v (t) \sim t^{-1.076}$.
If we assume that the vortices are uniformly (and randomly) distributed in 
two dimensional space, then using a naive argument, we would expect that   
$\rho_v (t) \sim L^{-2} (t)$. Interestingly, however, we find that the following nontrivial
scaling relation holds in the late time region (Fig.~3b)   
\begin{equation}
\rho_v (t) \sim  L^{-x}(t),    \qquad  x =1.93
\la{eq:rho_L}
\end{equation}
This feature is also very similar to the behavior of MCXY model and also that of the model E \cite{Nam_mcxy}. 
We note that the exponent $x$ showed little systematic variation at different values of $E$ except for some 
statistical numerical fluctuations (data not shown).                 

The nonequilibrium spin autocorrelation function $A (t)$ is expected to be related
to the growing length scale $L(t)$ through a new non-equilibrium exponent $\lambda$ as
\begin{equation}
A (t) \sim L^{-\lambda } (t).
\la{eq:A_L}
\end{equation}
We could extract the value of $\lambda$ by plotting $A(t)$ versus $L(t)$ as shown in
the Fig.~4, where we can see that, in the long time limit the value of 
$\lambda$ approaches $\lambda \simeq 2.21$ for the case of $E=0.4$. This value is rather close to 
the value in Ref. \cite{Nam_mcxy} (where $\lambda \simeq 1.99$ was observed) but much larger 
than the value of $\lambda \simeq 1.17 $ for the case of non-conserved $O(2)$ model with
no reversible mode coupling \cite{jrl, newman90,bh,newman90_2,lm92}.
It was argued in Ref.~\cite{Nam_mcxy} that the higher mobility of the vortices in the 
present model causes a faster memory loss of the initial configuration and hence larger value of 
the $\lambda$ exponent. It would be interesting if one can find some analytic 
argument for the value of $\lambda =2.21$. It appears that, due to the existence of propagating
spin wave modes, the spin configuration loses its memory more quickly than in the case of 
purely dissipative dynamics. We also found that there exists a very weak monotonic increase of the
$\lambda $ exponent with the increase of $E$ \cite{kknam_hxy_unpub}, which probably 
is related to the increasing spatial critical exponent $\eta (E)$ at finite temperature.
This trend is also consistent with the small difference between the value of $\lambda$ exponent in the
present work and that of MCXY or model E in Ref.~\cite{Nam_mcxy}, since, in the latter case, they are 
dealing with a zero temperature coarsening. 
Figure~5 shows a fitting of the full simulation results by our vortex-antivortex annihilation 
model via integration of Eq.~(\ref{eq:vort_anni_model}) with a suitable choice of the parameters. 



We note again that, in our Hamiltonian dynamics simulations, the initial
states are not completely random states due to the finite energy constraint,
Therefore, the initial states already have some finite correlation length 
scale, i.e., separation $R_0$ between vortices. These length scales are 
of the order of $3$ to $5$ lattice constants. Therefore, we expect that  
a finite time scale $\tau_0 $ (corresponding to the time scale for a vortex
to travel the distance $R_0$) exists after which a scaling region emerges.
It is interesting to note that the effect of the vortex inertia is observed 
to be not very significant at least for the domain growth law. This might 
be due to the finite initial length scale $L_0$. This should be contrasted to 
the case of the domain growth in the $O(2)$ Ginzburg Landau model with reversible 
mode coupling.    
The typical length scale $L_v(t=0)$ between vortices at initial time can be derived 
from the number density of vortices, which can be compared with the corresponding 
length scale $L_0$ derived from the equal-time spatial correlation functions. 
We expect that these two length scales are approximately proportional to each other,
which is indeed what we found from our data \cite{kknam_hxy_unpub}.
Note that $L_0$ was obtained by the arbitrary choice of $0.4$ for the value of 
the cut in the dynamic scaling collapse of the equal-time spatial correlation functions.  
Since $L_v (t=0)$ represents the initial average separation between vortices 
and antivortices at which the value of the rescaled correlation function would 
most probably be smaller than $0.4$, we expect that $L_v (t=0)$ is larger than $L_0$.
We could also confirm this from the data (data not shown here).


\section{Summary}

We have studied the coarsening dynamics of the Hamiltonian {\em XY}
model on two-dimensional square lattice. An initial state that is
specially tuned to have a given potential energy (otherwise random)
but with zero kinetic energy develops into a late-time coarsening
state, where the potential energy slowly decays as a power law with
a compensating increase in the kinetic energy. 
In order to compute accurately the equilibrium correlation exponents $\eta (E)$
for the order parameter, we employed an analog of the so-called NER 
method where the system begins with a perfect order but with finite kinetic energy. 
An interesting scaling behavior emerged here with linear growth of the
nonequilibrium correlation length. 
In the coarsening dynamics, the growth of the characteristic length scale 
can be fitted as $L(t) \sim  ((t+t_{0}) \ln(t+t_{0}))^{1/2}$ with a finite 
offset time $t_0$ which is due to the existence of a finite separation between
vortices and antivortices at the initial time.
The growth of the length scale can be interpreted as the vortex mobility 
diverging logarithmically in the size of the vortex-antivortex pair. 
A simple phenomenological dynamic model of vortex-antivortex annihilation 
can fit the growth of the length scale, with a suitable set of parameters for 
the effective inertia and mobility of vortices that are dependent on the 
growing length scale.
The nonequilibrium spin autocorrelation function $A(t)$ and the $L(t)$ 
are related through the nonequilibrium exponent $\lambda$ by $A(t) \simeq L^{-\lambda}(t)$ 
with an exponent of $\lambda \simeq 2.21$ (at $E=0.4$) which is distinctly different 
from the case of ordinary $O(2)$ or $XY$ model with dissipative dynamics.

\begin{acknowledgments}
This work was supported by Basic Science Research Program through 
the National Research Foundation of Korea (NRF) 
funded by the Ministry of Education, Science and Technology (No. 2010-0090085) (S. J. Lee) and
(No. 2012R1A2A2A01004172) (B. Kim and K. Nam)."
The computation of this work was supported by PLSI supercomputing resources of
Korea Institute of  Science and Technology Information (KISTI).

\end{acknowledgments}

\begin{appendix}
\section{ Critical dynamic scaling of the correlation functions in spin-wave approximation}
Here we analytically show that 
the critical dynamic scaling Eq. (7) holds in the spin-wave approximation.
We consider the equation of motion for the spin-angle variables $\theta_{i}$.
For convenience of notation, we denote here $\theta_{i}$ as $\theta ({\bf r},t)$
at position ${\bf r}$ (instead of $i$) and time $t$. 
In the spin-wave approximation with continuum notation, 
the Hamiltonian equation of motion reads as follows
\begin{equation} \label{eq:hxy-eq-spin_wave}
\frac{\partial^2 {\theta (\vec{r},t)}}{\partial t^2 } - \nabla^2 \theta (\vec{r},t)=0  
\end{equation}
which is the same as a wave equation. Here we simply put the wave velocity 
$c \equiv 1$.
We want to calculate the equal-time spatial correlation of the spins  for
the case of perfectly ordered  initial spin configuration (but with finite angular
velocities) which is defined as 
\begin{equation}
C_{nr} (r,t)  =  
\left < \exp \left [  i\theta ({\bf r}, t) - i \theta ({\bf 0},t) \right ] \right >,
\end{equation}
and the nonequilibrium spin autocorrelation function
\begin{equation}
A_{nr} (t) \equiv \left < \exp \left [ i \theta ({\bf r}, t) - i \theta ({\bf r},0) \right ] \right >.
\end{equation}
General solutions for the wave equation can be written as 
\begin{equation}
\theta ({\bf r},t) = \int_{{\bf k}} \left ( f^{+}_{\bf k} \exp ( i {\bf k} \cdot {\bf r}
 + i \omega t) +
                 f^{-}_{\bf k} \exp ( i {\bf k} \cdot {\bf r} - i \omega t) \right ) 
\end{equation}
where $\int_{\bf k} \equiv \int d^2 k /(2\pi)^2 $ and $\omega =|{\bf k}| = k $.
With the initial condition of $\theta ({\bf r}, t=0) = 0 $, we get
$ f^{+}_{\bf k} = -f^{-}_{\bf k} \equiv f_{\bf k} / 2i $ with  
\begin{equation}
\theta ({\bf r},t) = \int_{\bf k} f_{\bf k} \exp ( i {\bf k} \cdot {\bf r} )
 \sin (\omega t). 
\end{equation}
We assume that the initial angular velocity at each site has a Gaussian distribution
with zero mean and standard deviation proportional to the square root of the average energy per site.   
In addition, it is also assumed that the angular velocities exhibit delta-function correlations between different spatial points.  
If the average energy per site is given by $E$, then we have 
$\left < \dot{\theta} ({\bf r}, 0) \dot{\theta} ({\bf r}^{\prime}, 0) \right >  
= 2 E \delta ({\bf r}-{\bf r}^{\prime}) $, 
which, in the wave-vector space, can be written as 
\begin{equation}
\left < f_{\bf k} \,\, f_{{\bf k}^{\prime}} \right > =  \frac{8 \pi^{2} E}{k^2} 
\delta ({{\bf k}+{\bf k}^{\prime}}).
\end{equation}

Combining above relations and also using the gaussian nature, we
can obtain the following 
\begin{equation} \label{eq:equal-time_scaling}
\begin{array}{rcl}
C_{nr:SW}(r,t) 
&  = & 
\exp \left [ -\displaystyle\frac{1}{2} \left < \left ( \theta ({\bf r}, t) - \theta ({\bf 0},t) \right )^2 \right >   \right ] \\
      &     &        \\
      & =  & \exp \left [ -\displaystyle\frac{E}{\pi^2} \int d^2 k  \, \, \displaystyle\frac{1}{k^2} 
\sin^2 \big(\frac{1}{2} {\bf k} \cdot {\bf r} \big) \sin^2 (k t) \right ]. 
\end{array}
\end{equation}
In order to compute the integral inside the exponential, we put 
\begin{equation}
J(\vec{r}, t) \equiv \int d^2 k  \, \, \frac{1}{k^2} 
\sin^2 \big(\frac{1}{2} {\bf k} \cdot {\bf r} \big) \sin^2 (k t)
\end{equation}
Taking a derivative of $J$ with respect to $t$ and also adding prescription for the proper convergence
in the (positive) large $k$ limit, we arrive at  
\begin{equation}
\begin{array}{rcl}
\displaystyle\frac{\partial J({r}, t)}{\partial t} & = & \lim_{\epsilon \rightarrow +0} 
\frac{1}{2} \int_{0}^{2 \pi} d\phi \, \int_{0}^{\infty} d k\, \sin (2 k t) \\
       &     &   \times \Big( 1-\cos \big( k \, r \cos \phi \big) \Big) \, \exp(-\epsilon k).
\end{array}
\end{equation}
This integral can now be easily done with the following result
\begin{equation} \label{eq:dJ_over_dt}
\begin{array}{rcl}
\displaystyle\frac{\partial J({r},t) }{\partial t} & =  & \displaystyle\frac{\pi}{2t} 
\displaystyle\left ( 1 - \big[1-(r/2t)^2\big]^{-1/2} \right ) 
      \hspace{.5cm} (\frac{r}{2t} < 1),  \\
            &      &            \\
        &  =  & {\displaystyle\frac{\pi}{2t}} \hspace{1.0cm} (\displaystyle\frac{r}{2t} > 1). 
\end{array}
\end{equation}
The above equation can now be integrated with respect to $t$ with proper boundary conditions, leading to 
the equal-time spatial correlation function as
\begin{equation} \label{eq:NER_scaling_spin_wave}
C_{nr:SW}(r,t) \simeq r^{-\eta_{SW}(E)} g_{nr:SW}(r/2t)
\end{equation}
where $\eta_{SW} (E) = E / 2 \pi $ and
\begin{equation} \label{eq:scaling_f_spin_wave}
\begin{array}{rcl}
g_{nr:SW}(u) & =  &  u^{\eta_{SW} (E)} 
\left (\displaystyle\frac{1 + \sqrt{1-u^2}}{ 1 - \sqrt{1-u^2}} \right )^{\eta_{SW} (E)/2 } \hspace{0.2cm} (u<1), \\
      &      &          \\
     & = &  u^{\eta_{SW} (E)} \hspace{0.5cm} (u > 1). 
\end{array}
\end{equation}
In the limit of $u \equiv r/2t << 1$, we get $C_{nr:SW} (r,t) \sim r^{-\eta_{SW}(E)} (1- \eta_{SW}(E) u^2 /8 )$
and also in the other regime of $ u \equiv r/2t > 1 $,  we get $C_{nr:SW} (r,t) \sim t^{-\eta_{SW}(E)} $ 
independent of the value of $r$. This agrees qualitatively with the simulation results including 
the plateau behavior in the region of $ u \equiv r/2t > 1 $. The value of $\eta_{SW} (E)$ at spin-wave approximation for $E=0.4$ is 
given by $E/2 \pi \simeq 0.064 $ which is in approximate agreement with the simulation result of 
$\eta (E=0.4) = 0.071$ (see Fig.~1c). 
Also we find for the autocorrelation $A_{nr:SW}(t)$
\begin{equation} \label{eq:autocorr_spin_wave}
A_{nr:SW}(t) \simeq t^{-\eta_{SW} (E)/2} 
\end{equation}
which again shows approximate agreement with simulations results (data not shown).

\end{appendix}

\newpage

\begin{figure}[t]
\includegraphics[angle=0,width=8cm]{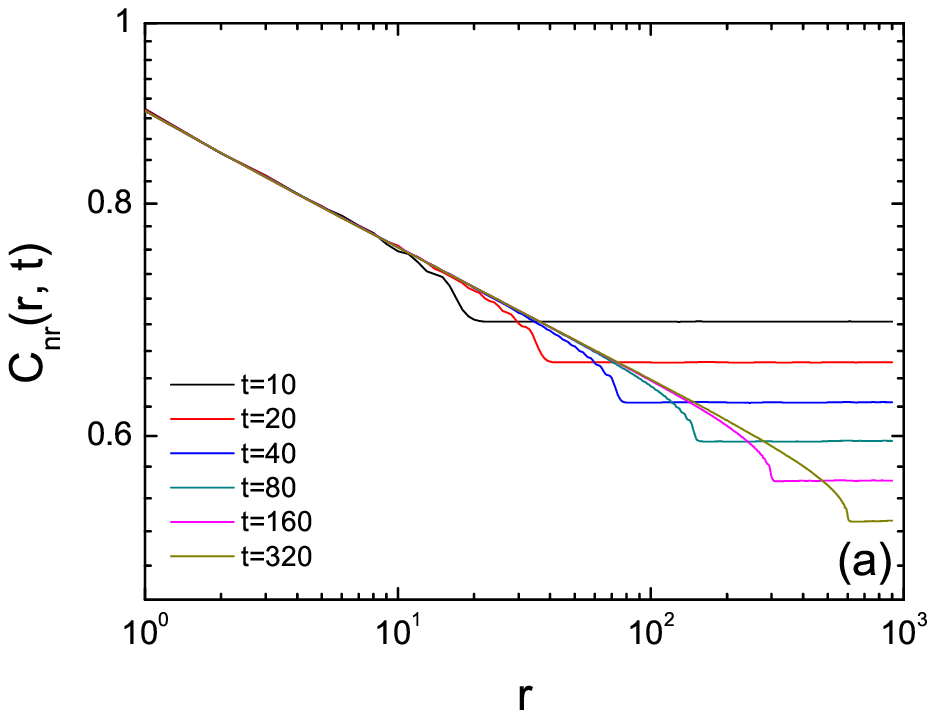}
\includegraphics[angle=0,width=8cm]{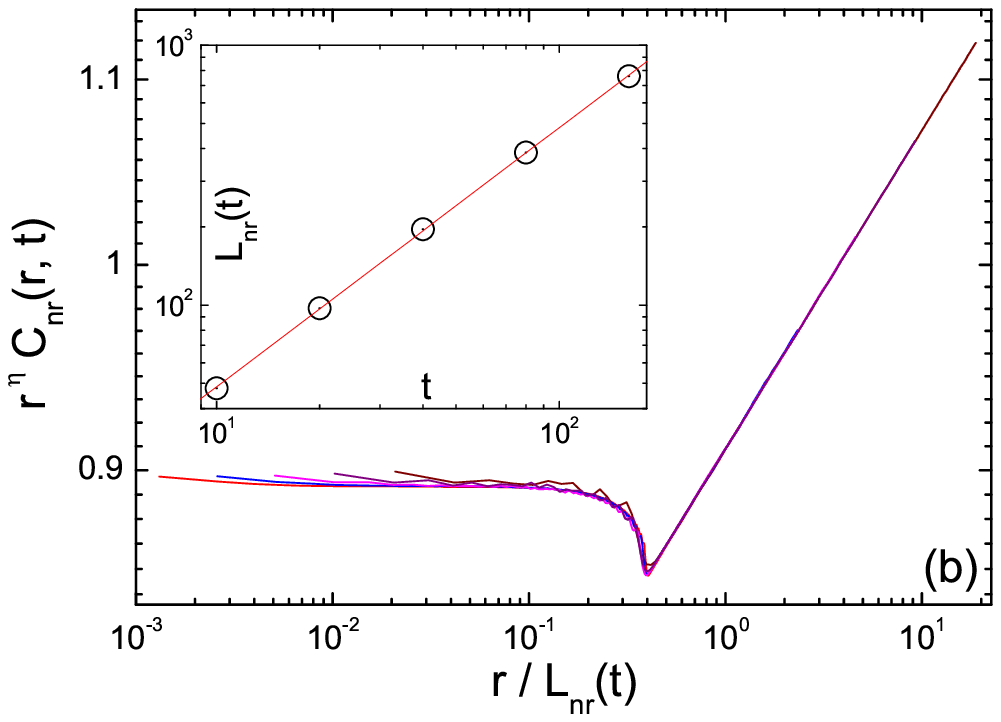}
\includegraphics[angle=0,width=8cm]{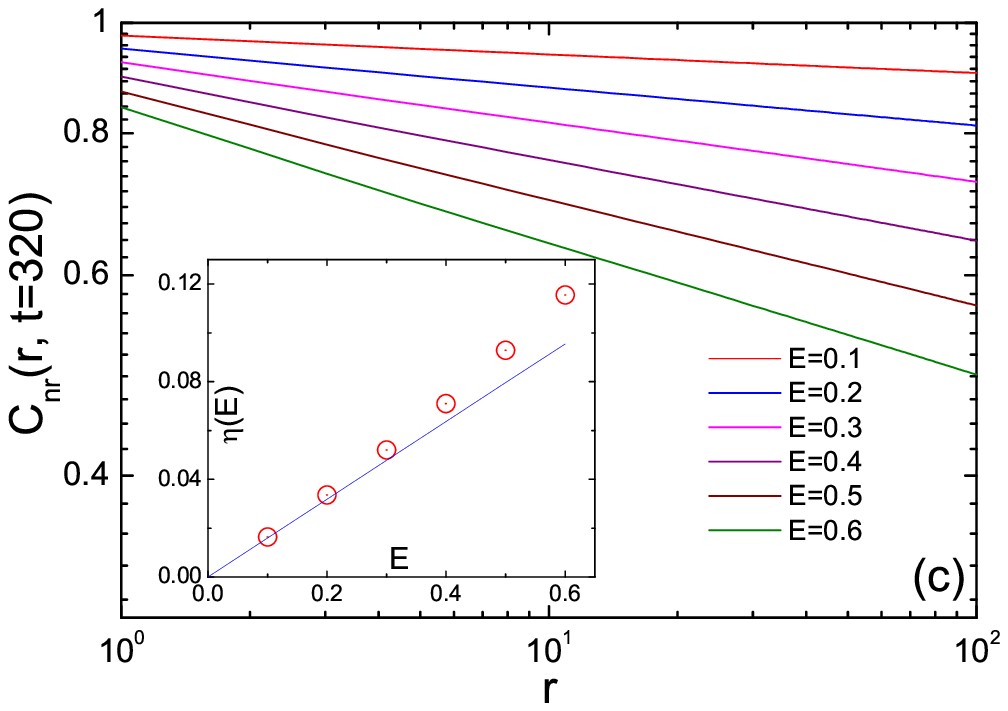}
\caption{
(a) The equal-time spatial correlation function of the $XY$ spin order 
parameter with NER dynamics for $E=0.4$ at times $t=10$, $20$, $40$, $80$, 
$160$, $320$, where the system size is $1800 \times 1800$.  
(b) The scaling collapse of the data in (a) with the appropriate scaling 
length $L_{nr}(t)$. The inset shows the length scale $L_{nr} (t)$ vs. $t$
which exhibits a linear growth $L_{nr} (t) \sim t^{1/z}$ with $1/z \simeq 1 $
and $\eta \simeq 0.071 $.   
(c) The equal-time spatial correlations $C_{nr}(r,t)$ at late time $t = 320$ for
  different energies. The inset shows the corresponding $\eta (E)$ (circles) and 
the spin-wave approximation values $\eta_{SW}(E)$ (solid line) versus $E$.
} \label{fig1}
\end{figure}

\begin{figure}[t]
\includegraphics[angle=0,width=8cm]{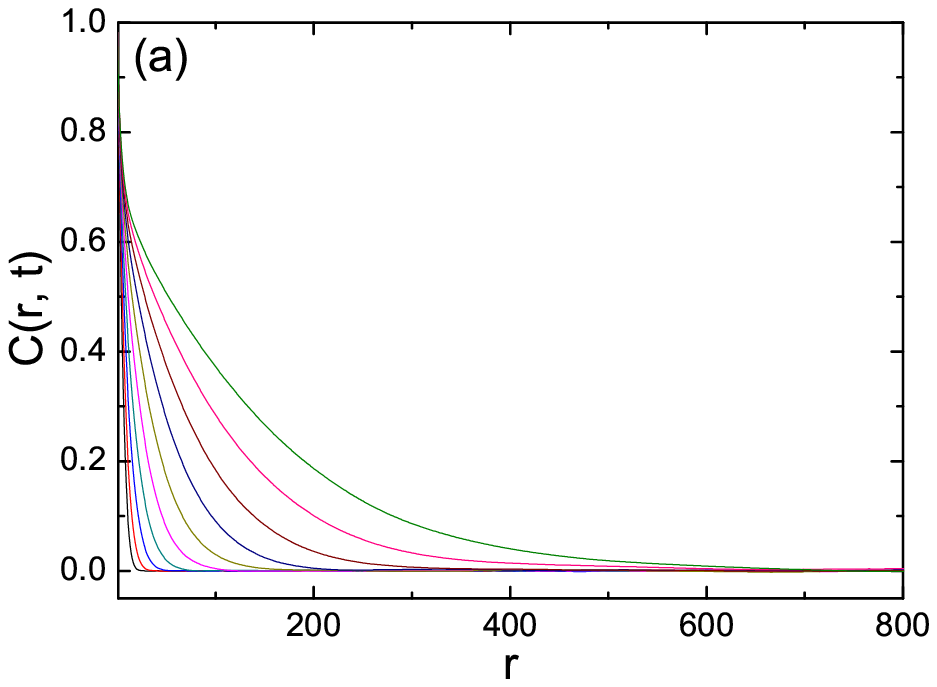}
\includegraphics[angle=0,width=8cm]{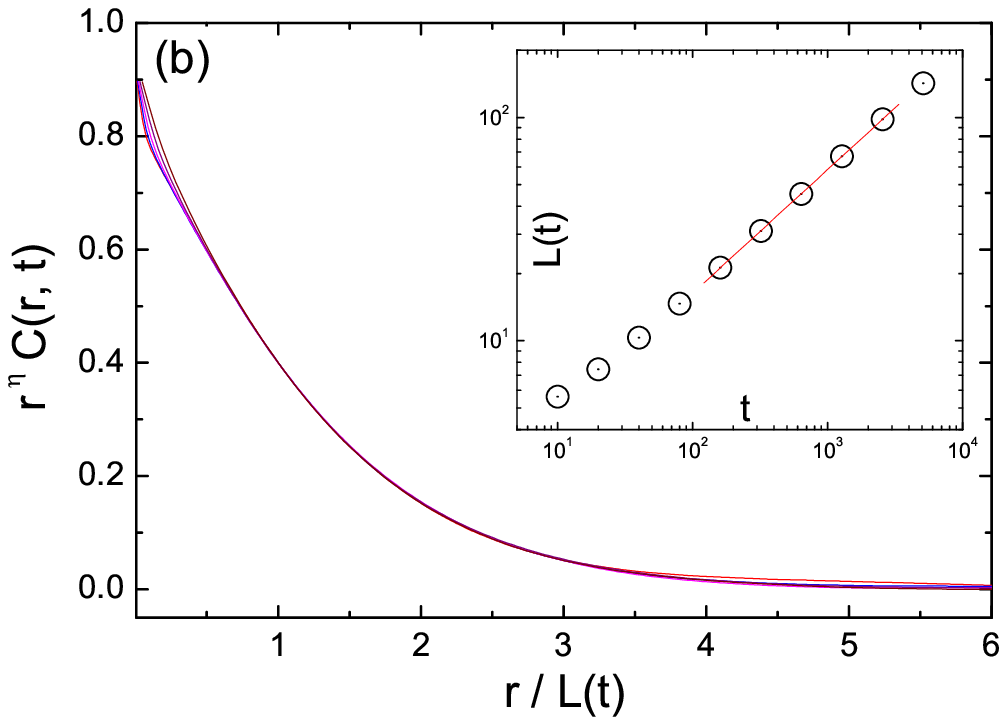}
\includegraphics[angle=0,width=8cm]{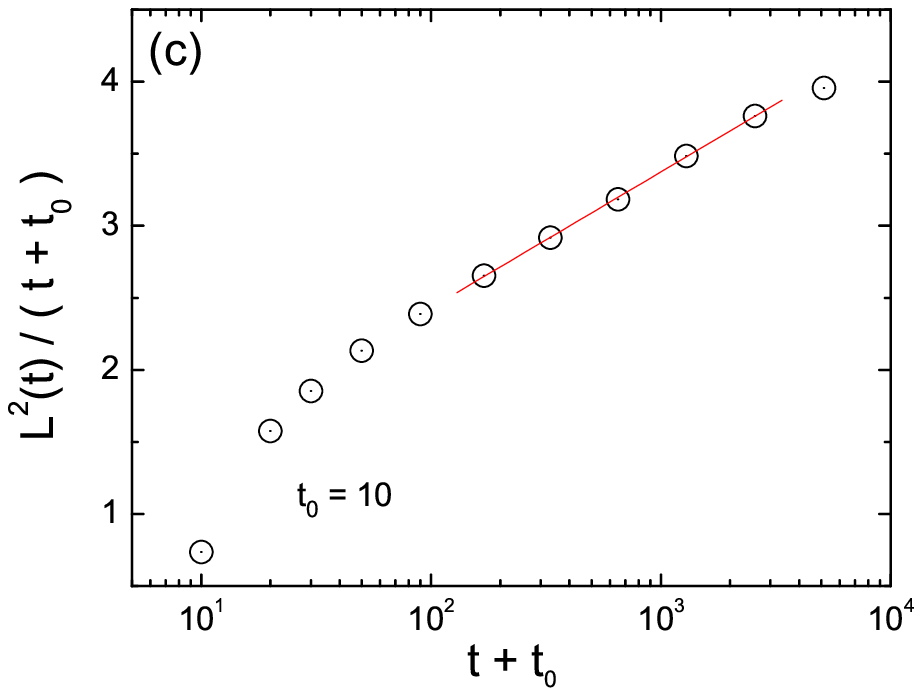}
\caption{
(a) The equal-time spatial correlation function of the $XY$ spin order 
parameter with coarsening dynamics for $E=0.4$ at various times from $t=10$ 
up to $5120$ with the system size is $2000 \times 2000$.  
(b) The scaling collapse of the data in (a) with the appropriate scaling 
length $L(t)$ for $t=160$, $320$, $640$, $1280$ and $2560$. The inset shows 
the length scale $L (t)$ vs. $t$ which exhibits an effective superdiffusive 
power law growth of $L(t) \sim t^{1/z}$ with $1/z \simeq 0.553$.   
(c) $L^2 (t)/(t + t_0 ) $ vs. $\ln (t + t_0 )$ which indicates a positive
logarithmic correction for the growing length scale.        
} \label{fig2}
\end{figure}

\begin{figure}[t]
\includegraphics[angle=0,width=8cm]{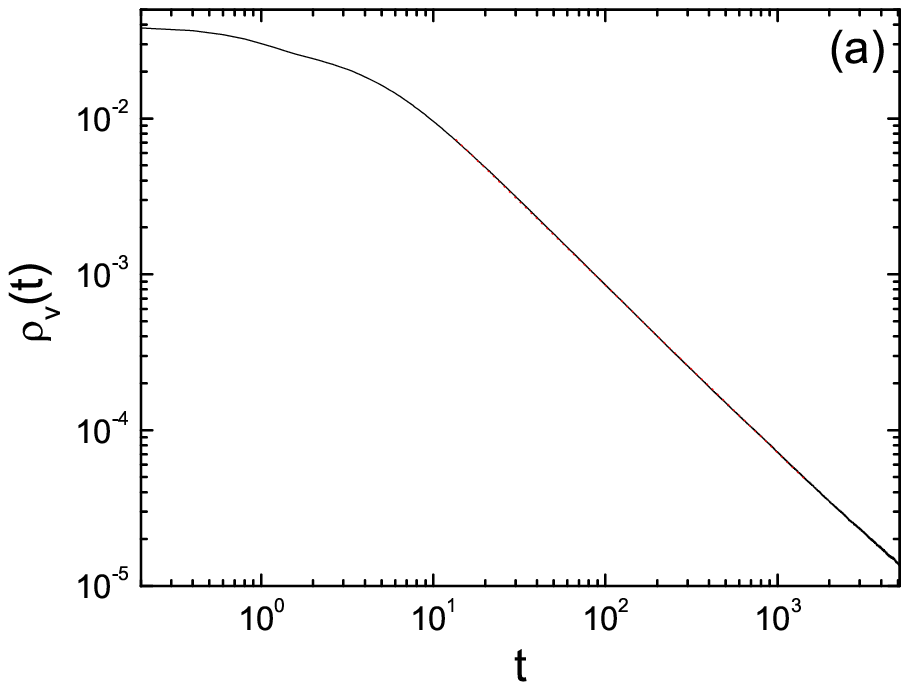}
\includegraphics[angle=0,width=8cm]{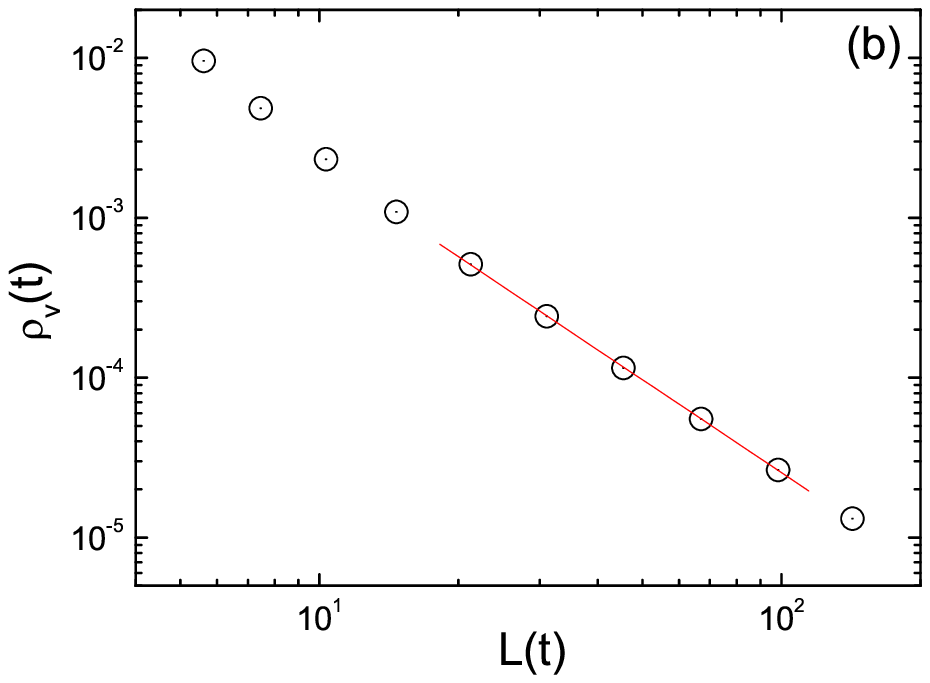}
\caption{
Relaxation of (a) the vortex number density $\rho_v (t)$ at energy 
$E=0.4$ and system size $2000 \times 2000$ which shows a power law decay
with $\rho_v (t) \sim t^{-1.076}$ (in the time interval between $t=10$ and $1000$).
(b) The vortex number density  $\rho_v(t)$  vs. $L(t)$ exhibits a power law decay
$\rho_v(t) \sim L^{-x}(t) $ with $x \simeq 1.93 $.
} \label{fig3}
\end{figure}

\begin{figure}[t]
\includegraphics[angle=0,width=8cm]{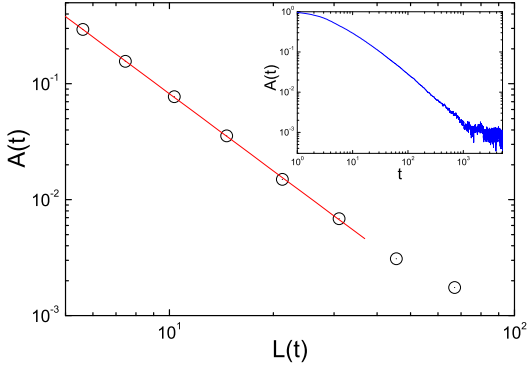}
\caption{
Spin autocorrelation function vs. $L(t)$ at energy $E = 0.4$ 
indicating a simple power law relationship between the two with 
$A(t) \sim L(t)^{-2.21}$. 
} \label{fig4}
\end{figure}

\begin{figure}[t]
\includegraphics[angle=0,width=8cm]{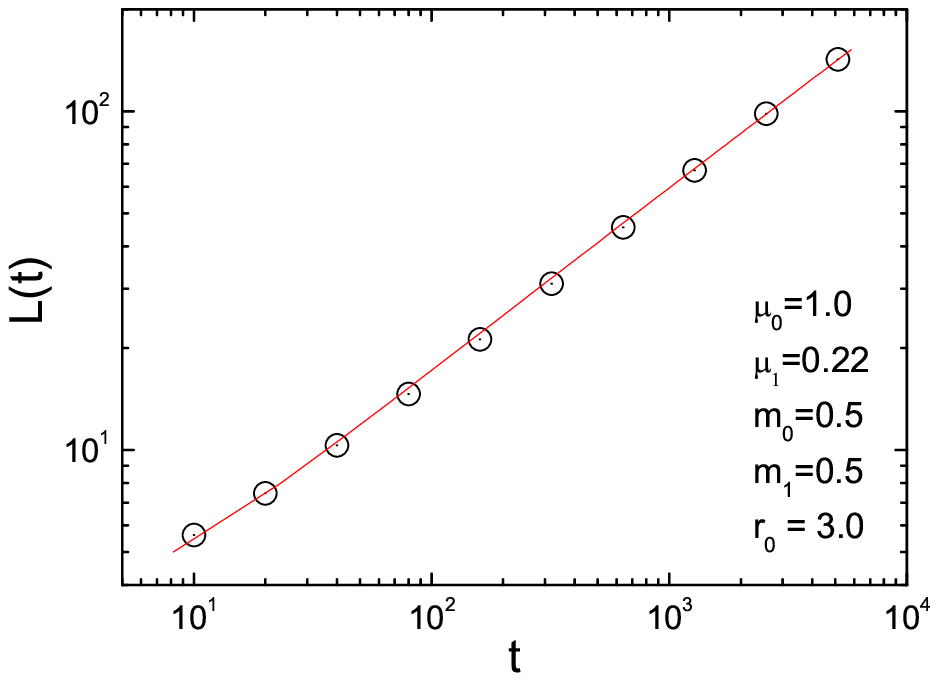}
\caption{
The growth of $L(t)$ fitted by a simple vortex-antivortex annihilation model
at energy $E = 0.4$ which shows a good agreement. 
} \label{fig5}
\end{figure}

%
%

%
%


\begin{references}

\bibitem{ordering_review} J. D. Gunton, M. San Miguel, and P. S. Sahni, in
 {\it Phase Transitions and Critical Phenomena}, edited by
  C. Domb and J. L. Lebowitz (Academic, New York, 1983), Vol 8;
  H. Furukawa, Adv. Phys. {\bf 34}, 703 (1985); K. Binder, Rep. Prog.
  Theor. Phys. {\bf 50}, 783 (1987).


\bibitem{bray_review}
A. J. Bray, Adv.\ Phys.\ {\bf 43}, 357 (1994).

\bibitem{ma_mazenko} S. Ma and G. F. Mazenko, Phys. Rev. B {\bf 11}, 4077 (1975).

\bibitem{nelson-fisher}
D. R. Nelson and D. S. Fisher, Phys. \ Rev. \ B {\bf 16}, 4945 (1977).

\bibitem{Nam_mcxy}
K. Nam, B. Kim, and S. J. Lee, J. Stat. Mech. Theo. Exp., {\bf 2011}, P03013 (2011).

\bibitem{hohen77}
P. C. Hohenberg and B. I. Halperin, Rev. Mod. Phys. {\bf 49}, 435 (1977).

\bibitem{folk2006} R. Folk and G. Moser, J. Phys. A: Math. Gen. {\bf 39}, R207 (2006).

\bibitem{casetti_hamil_review}
L. Casetti, M. Pettini, and E. G. D. Cohen, \ Phys. \  Rep.\ {\bf 337} 237 (2000).

\bibitem{caian1}
L. Caiani, L. Casetti, C. Clementi, G. Pettini, M. Pettini, and R.
Gatto, \ Phys. \ Rev. \ E {\bf 57}, 3886 (1998).

\bibitem{caian2}
 L. Caiani, L. Casetti, and M. Pettini, \ J. \ Phys. \ A \ {\bf 31}, 3357 (1998).

\bibitem{leoncini}
 X. Leoncini, A. D. Verga, and S. Ruffo, \ Phys. \ Rev. \ E {\bf 57}, 6377
 (1998).

\bibitem{ruffo2001}
 S. Lepri, and S. Ruffo, \ Europhys. \ Lett. {\bf 55}, 512 (2001).

\bibitem{cerruti}
M. Cerruti-Sola, C. Clementi, and M. Pettini, \ Phys. \ Rev. \ E
{\bf 61}, 5171 (2000).

\bibitem{latora}
V. Latora, A. Rapisarda, and S. Ruffo,  \ Physica \ D \ {\bf 131}, 38
(1999).

\bibitem{zheng_ising_ordering}
 B. Zheng, \ Phys. \ Rev. \ E {\bf 61}, 153 (2000).

\bibitem{kockel_ising_ordering_comment}
J. Kockelkoren and H. Chat\'{e}, Phys.\ Rev.\ E {\bf 65}, 058101
(2002).

\bibitem{koo2006}
K. Koo, W. Baek, B. Kim, and  S. J. Lee, J.\ Korean\ Phys.\ Soc. {\bf 49}, 1977 (2006).

\bibitem{BKT} V.\ L.\ Berezinskii, Zh.\ Eksp.\ Teor.\ Fiz.\ {\bf 59}, 907 (1970)
[Sov.\ Phys.\ JETP {\bf 32}, 493 (1971)]; J.\ M.\ Kosterlitz, and D.\ J.\
Thouless, J.\ Phys.\ C {\bf 6}, 1181 (1973); J. \ M.\ Kosterlitz, {\it
ibid.} {\bf 7}, 1046 (1974).






\bibitem{loft}
R. Loft and T. A. DeGrand, \ Phys.\ Rev. \ B {\bf 35}, 8528 (1987);
H. Toyoki and  K. Honda, \ Prog.\ Theor. \ Phys. {\bf 78}, 237 (1987).


\bibitem{toyoki3} H. Toyoki, Phys. Rev. A {\bf 42}, 911 (1990).


\bibitem{mondello}
M. Mondello and  N. Goldenfeld, \ Phys. \ Rev. \ A {\bf 42}, 5865 (1990);
Phys. Rev. E {\bf 47}, 2384 (1993).

\bibitem{bray}
A. J. Bray and K. Humayun, J. \ Phys. \ A {\bf 23}, 5897 (1990);
S. Puri and C. Roland, \ Phys. \ Lett. \ A {\bf 151}, 500 (1990);
H. Toyoki, \ Phys. \ Rev. \ A  {\bf 42}, 911 (1990).

\bibitem{yurke}
B. Yurke, A. N. Pargellis, T. Kovacs, and D. A. Huse, \ Phys. \
Rev. \ E {\bf 47}, 1525 (1993).


\bibitem{blundell}  R. E. Blundell and A. J. Bray, Phys. Rev. E {\bf 49},
            4925 (1994).

\bibitem{jrl}
J-R. Lee, S. J. Lee, and B. Kim, {\it ibid}. {\bf 52}, 1550 (1995).

\bibitem{rojas}
F. Rojas and A. D. Rutenberg, \ Phys. \ Rev. \ E {\bf 60}, 212
(1999).

\bibitem{bray2}
A. J. Bray, A. J. Briant, and D. K. Jervis, \ Phys. \ Rev. \ Lett.
{\bf 84}, 1503 (2000);

\bibitem{bray3}
A. J. Bray, \ Phys. \ Rev. \ E {\bf 62}, 103 (2000).

\bibitem{ying}
H. P. Ying, B. Zheng, Y. Yu, and S. Trimper, \ Phys. \ Rev. \ E {\bf
63}, 035101 (2001).


\bibitem{wilczek} A. A. Ovchinnikov and Ya. B. Zeldovich, Chem. Phys.
     {\bf 28}, 214 (1978); D. Toussaint and F. Wilczek, J. Chem.
     Phys. {\bf 78}, 2642 (1983); K. Kang and S. Redner,
     Phys. Rev. A {\bf 30}, 2833 (1984).

\bibitem{jj_ordering}
G. S. Jeon, S. J. Lee, and M. Y. Choi, Phys. \ Rev. \ B {\bf 67},
014501 (2003).




\bibitem{newman90} T. J. Newman and A. J. Bray, J. Phys. A {\bf 23}, L279 (1990);
{\bf 23}, 4491 (1990).

\bibitem{bh}
A. J. Bray and K. Humayun, J. \ Phys. \ A {\bf 23}, 5897 (1990);

\bibitem{newman90_2} T. J. Newman, A. J. Bray, M. A. Moore, Phys. Rev. B {\bf 42}, 4514 (1990).

\bibitem{lm92} F. Liu and G. F. Mazenko, \ Phys. \ Rev. \ B {\bf 45}, 6989 (1992);
{\bf 46}, 5963 (1992).

\bibitem{das-rao1}
J. Das and M. Rao, \ Phys. \ Rev. \ E {\bf 57}, 5069 (1998).

\bibitem{das-rao2}
J. Das and M. Rao, \ Phys. \ Rev. \ E {\bf 62}, 1601 (2000).



\bibitem{yoshida}  H. Yoshida, Phys. \ Lett. A {\bf 150}, 262 (1990).

\bibitem{kknam_hxy_unpub}
K. Nam, S. J. Lee, and B. Kim, unpublished.

\bibitem{sane_2009} J. San\'{e}, J. T. Padding, A. A. Louis, Phys. Rev. E {\bf 79}, 051402 (2009).

\bibitem{saffman_1975} P. G. Saffman and M. Delbr\"{u}ck, Proc. Nat. Acad. Sci. {\bf 72}, 3111 (1975).

\bibitem{bishop}
F. G. Mertens and A. R. Bishop, {\em Dynamics of Vortices in
Two-Dimensional Magnets}, editted by P. \ L. Christiansen and M. \ P. Sorensen,
 {\em Nonlinear Science at the Dawn of the 21st Century},
(Springer, Berlin, 1999).

\bibitem{kamppeter1}
T. Kamppeter, F. G. Mertens, E. Moro, A. S\'{a}nchez, and A. R. Bishop,
 \ Phys. \ Rev. \ B {\bf 59}, 11349 (1999).

\bibitem{kamppeter2}
T. Kamppeter, F. G. Mertens, A. S\'{a}nchez, A. R. Bishop, F.
Dom\'{i}nguez-Adame, and N. Gr{$\phi$}nbech-Jensen, \ Eur. \ Phys. \ J.
\ B {\bf 7}, 607 (1999).



\end{references}
\end{document}